\documentstyle[preprint,aps,amssymb,psfig]{revtex}
\begin{document}
\tighten
\draft


\title{The dynamics of sodium in sodium disilicate: Channel relaxation 
       and sodium diffusion}                           
\author{J\"urgen Horbach$^{{\rm (1)}}$, 
        Walter Kob$^{{\rm (2)}}$, and 
        Kurt Binder$^{{\rm (1)}}$}

\address{$^{{\rm (1)}}$Institut f\"ur Physik, Johannes Gutenberg--Universit\"at Mainz\\
         Staudinger Weg 7, D--55099 Mainz, Germany\\
         $^{{\rm (2)}}$Laboratoire des Verres - Universit\'e Montpellier 2\\
         Place E. Bataillon, cc 069, 34095 Montpellier, France}

\date{\today}

\maketitle


\begin{abstract}
We use molecular dynamics computer simulations to study the dynamics of
amorphous (Na$_2$O)2(SiO$_2$). We find that the Na ions move in channels
embedded in a SiO$_2$ matrix. The characteristic distance between these
channels gives rise to a prepeak in the structure factor at around
$q=0.95$~\AA$^{-1}$. The dynamics of sodium is given by a fast
process which can be seen in the incoherent scattering function and
a slow process which is seen in the coherent function. The relaxation
time of the latter coincides with the $\alpha-$relaxation time of the
matrix. The Kohlrausch exponent of the fast process for $q>1.6$~\AA$^{-1}$
is the same as the von Schweidler exponent for the slow one, demonstrating
that the two processes are closely related.

\end{abstract}



The dynamics of mobile ions in glasses and melts is a field of
great interest in geoscience, materials science, chemistry, and
physics~\cite{rev_id,maass92}. However, due to the lack of experiments
that probe the dynamics on a microscopic level, many details of this
dynamics are not understood well. A very valuable tool to understand
these details are molecular dynamics (MD) computer simulations. Many
of the recent MD studies of alkali silicate glasses~\cite{mdalk} have
given support to the idea of Ingram~\cite{ingram89} that alkali ions move
through ``preferential pathways'' and that these pathways, as has been
suggested by Greaves~\cite{greaves85}, are related to a microsegregation
of alkali ions. Very recently Jund {\it et al.}~\cite{jund01} investigated
the dynamics of the Na ions in sodium tetrasilicate (NS4) and found {\it no}
evidence for a clustering of sodium ions in individual configuration
snapshots.  However, these authors did identify preferential pathways
in the dynamics of the sodium ions in that the sodium trajectories form
a well connected network of pockets and connecting pathways. Moreover, they gave
evidence for an uncorrelated motion of the sodium ions inside these
channels, contrarily to the picture of a cooperative motion proposed
by Greaves~\cite{greaves85}.

What so far has been left out from the discussion is the slow relaxation
dynamics of the matrix, i.e.~in sodium silicates the dynamics of
the silicon and oxygen atoms. This is a very important issue since
most of the models for the ion transport in glasses start from a
coarse--grained representation of a glassy matrix which is not derived
from first principles~\cite{maass92}. Thus the present paper is the first
investigation in which the dynamics of the fast sodium ions is compared
to the slow relaxation of the matrix and below we will show that these
two processes are much closer related to each other than could naively
be expected.

The system studied is (Na$_2$O)2(SiO$_2$), NS2, i.e. a prototype of an
ion conducting glass. The potential we use to describe the interactions
between the ions in our MD simulation is a slight modification of the pair
potential proposed by Kramer {\it et al.} which is based on {\it ab
initio} calculations~\cite{kramer91}. More details on this potential can
be found in Ref.~\cite{horbach01}. The simulation was done at constant
volume using 8064 particles ($N_{{\rm Si}}=1792$, $N_{{\rm Na}}=1792$,
and $N_{{\rm O}}=4480$) in a box of size $L=48.$653~\AA. This corresponds
to a mass density of 2.37~g$/$cm$^3$, which is close to the experimental
one~\cite{mazurin}.  The system was equilibrated at various temperatures
in the range 2100~K~$\le T \le$~4000~K before the production runs were
started. These were done in the microcanonical ensemble, using the
velocity form of the Verlet algorithm with a step size of 1.6~fs. At the
lowest temperatures, the temperature for which the results presented here
were obtained, these runs extended over 2.5~ns (1.5 million time steps)
and in order to improve the statistics of the results we averaged over
two independent runs.

In a recent publication we have shown that this model predicts structural
and dynamical properties of NS2 which are in good agreement with
experimental findings~\cite{horbach01}. In particular we found that
the static structure factor exhibits a prepeak at $q=0.95$~\AA$^{-1}$
which corresponds to a length scale of next-nearest Na-Na or Si--Na
neighbors (around 6.6~\AA). This result was very recently confirmed
in a neutron scattering experiment by Meyer {\it et al.}~\cite{meyer01}
who found at this wave--vector a pronounced shoulder in the elastic
intensities. Furthermore it has been shown that at temperatures $T \le
2500$~K the dynamics of the Na atoms is about two orders of magnitude
faster than the one of silicon and oxygen atoms, which is in qualitative
agreement with the experimental fact that this system is an ion conducting
material~\cite{horbach01}.  Hence it is reasonable to assume that the present model does
indeed give a good microscopic description of real NS2.

\begin{figure}
\psfig{file=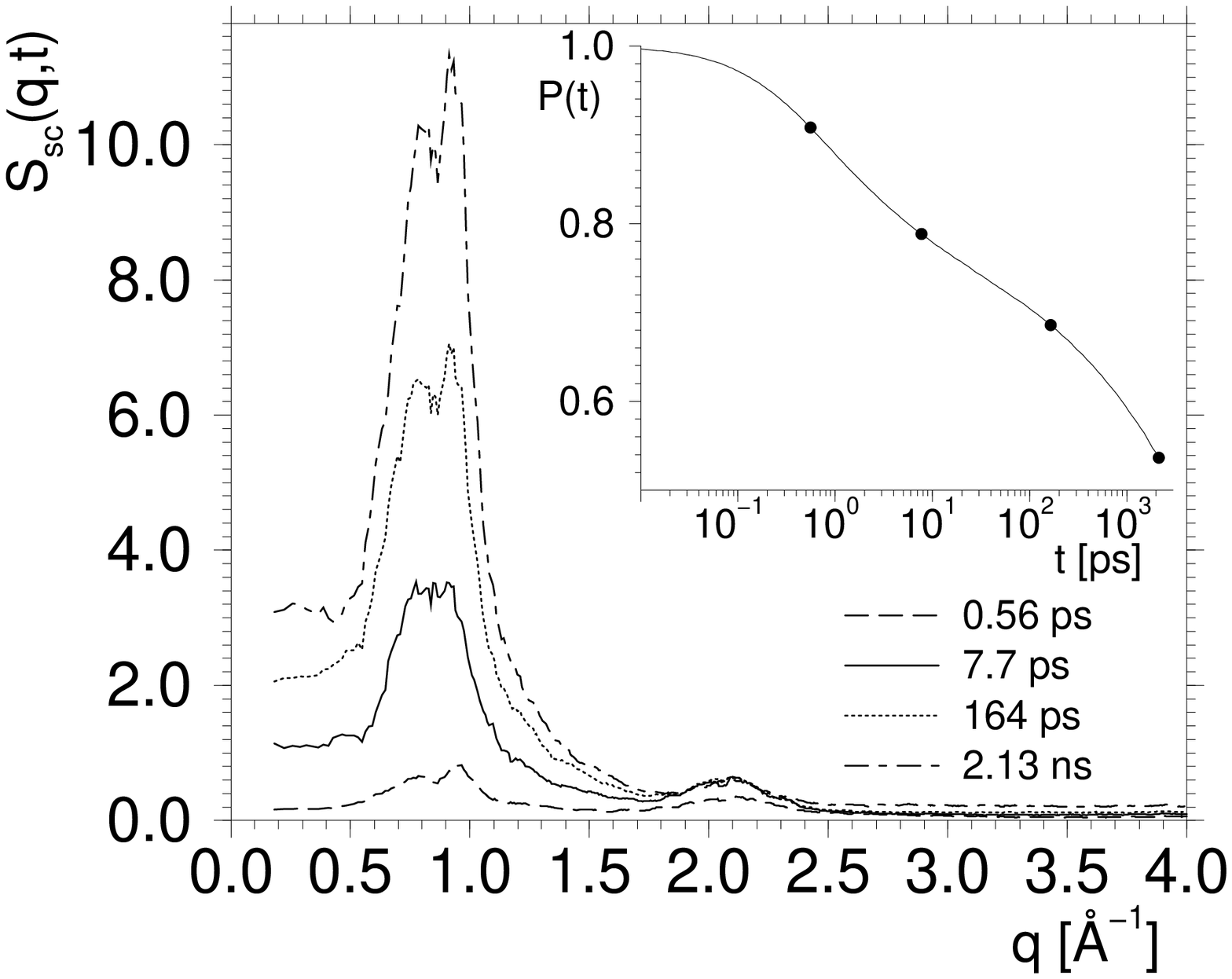,width=10cm,height=8.5cm}
\vspace{0.2cm}
\caption{``Swiss cheese'' structure factor $S_{\rm sc}(q,t)$ for the sodium free regions
         at $T=2100$~K for $t=0.56$~ps, 7.7~ps, $164$~ps, and $2.13$~ns. The inset shows the
	 probability $P(t)$ that a cube which is sodium free at time zero remains sodium free until time
	 $t$. The circles on the curve for $P(t)$ are at the times at which 
	 $S_{\rm sc}(q,t)$ is shown.}
\label{fig1}
\end{figure}

Using the same type of potential for NS4 it was recently shown
that the Na atoms move in preferential pathways, also called
``channels'', which are basically frozen in structures inside the
SiO$_2$ matrix~\cite{jund01}. Furthermore it was found that the volume
of these channels is relatively low and that they contain a large
number of non-bridging oxygen atoms~\cite{ispas01,sunyer01}.  Hence it can be
expected that the average density of the sodium atoms {\it outside}
these channels is significantly lower than can be expected if the
sodium atoms were distributed inside the system in a uniform way,
i.e. that the diffusion of the sodium atoms is strongly restricted to
a small subset of the configuration space. In order to check this idea
we have calculated a (coarse grained) probability of finding {\it no}
sodium atom at a given location in space. Following the approach of Jund
{\it et al.}~\cite{jund01} we calculated this probability by dividing the
system into $48^3$ cubes (of length $L/48 \approx 1.01$~\AA). Then we calculate
the probability $P(t)$ that a cube which does not contain a sodium ion at
time zero is also not visited by a sodium ion until a later time $t$. The
time dependence of $P(t)$ is shown in the inset of Fig.~\ref{fig1}. From
this graph we recognize that after 2.5~ns, i.e. after more than the
$\alpha-$relaxation time of the matrix~\cite{horbach01a}, more than 50\%
of the cubes have not yet been visited by a sodium atom. (We mention
that after this time the mean squared displacement of the Na atoms is
more than (45~\AA)$^2$, which shows that these atoms have moved a large
distance. On this time scale also the local structure of the Si--O matrix
is reconstructed~\cite{horbach01a}.)  Hence we can conclude that on this time
scale the sodium free region forms a percolating structure
around a network of channels, i.e.~it has somewhat the structure of a
Swiss cheese.  In order to investigate the structure of this percolating
region we define a ``Swiss cheese'' structure factor $S_{\rm sc}(q,t)$ 
as follows: We assign to each cube which has not been visited by
a sodium atom until time $t$ a point and we compute the static structure
factor of $N_{{\rm sc}}(t)=P(t) (48^3-N_{{\rm Na}})$
points:

\begin{equation}
   S_{\rm sc}(q,t)= \frac{1}{N_{{\rm sc}}(t)} \sum_{k,l=1}^{N_{{\rm sc}}(t)} 
       \exp(i \vec{q} \cdot (\vec{r}_k - \vec{r}_l)) \ . 
\end{equation}
\label{eq1}
This quantity is shown in Fig.~\ref{fig1} for four different times:
$t=0.56$~ps, $7.7$~ps, $164$~ps, and $2.13$~ns. 
We see that $S_{\rm sc}(q,t)$ has peaks at $q_1=0.9$~\AA$^{-1}$
and $q_2=2.15$~\AA$^{-1}$ which are also the prominent features in
$S_{\rm NaNa}(q)$, the static structure factor for the Na--Na
correlations~\cite{horbach01}. Hence we can now conclude that the peak
at $q_1$ in $S_{\rm NaNa}(q)$, which has also been seen in n-scattering
experiments~\cite{meyer01}, corresponds to the typical distance between
the channels. Note that with increasing time the height of this peak
increases quickly. However, it is clear that the peak at $q_1$
decreases again on the time scale on which the matrix starts to reconstruct itself
significantly and thus rearranges the channel structure.

The peak at $q_2$ reflects the length scale of nearest sodium neighbors
\cite{horbach01}. As we see in Fig.~\ref{fig1} its height is constant
for $7.7$~ps$ \le t \le 2.13$~ns. The fact that this peak is still present
at $t=2.13$~ns means that there are preferred sites for the sodium ions
inside the channels which is reasonable because most of the sodium ions
are located in the neighborhood of dangling bonds~\cite{ispas01,sunyer01}
and these bonds do not disappear until the network regions have
significantly rearranged themselves. The latter finding is also in
agreement with the results from Ref.~\cite{jund01} for the distinct part
of the van Hove correlation function.

\begin{figure}
\psfig{file=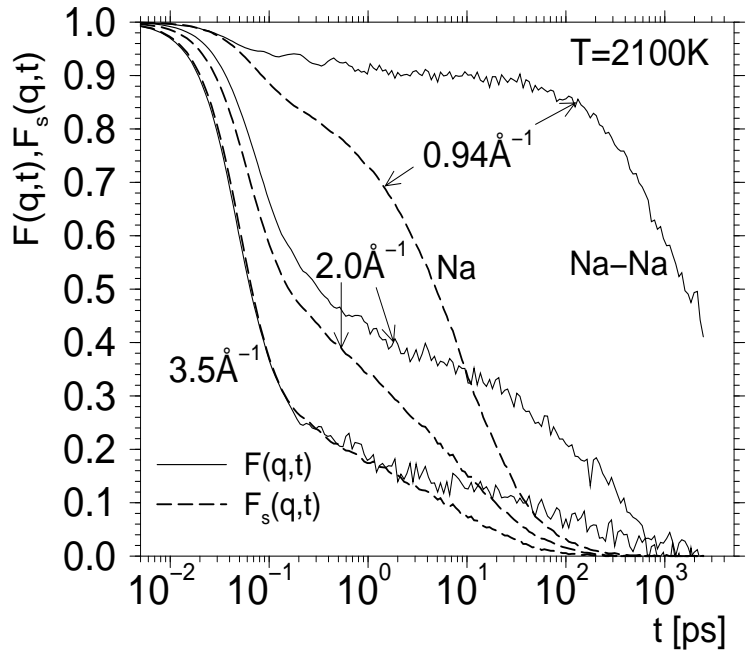,width=10cm,height=8.5cm}
\vspace{0.2cm}
\caption{Coherent intermediate scattering functions $F(q,t)$ for the sodium--sodium 
         correlations (solid lines) and incoherent intermediate scattering functions 
	 $F_{{\rm s}}(q,t)$ (dashed lines) at $T=2100$~K for the indicated values of
	 $q$.}
\label{fig2}
\end{figure}

\begin{figure}
\psfig{file=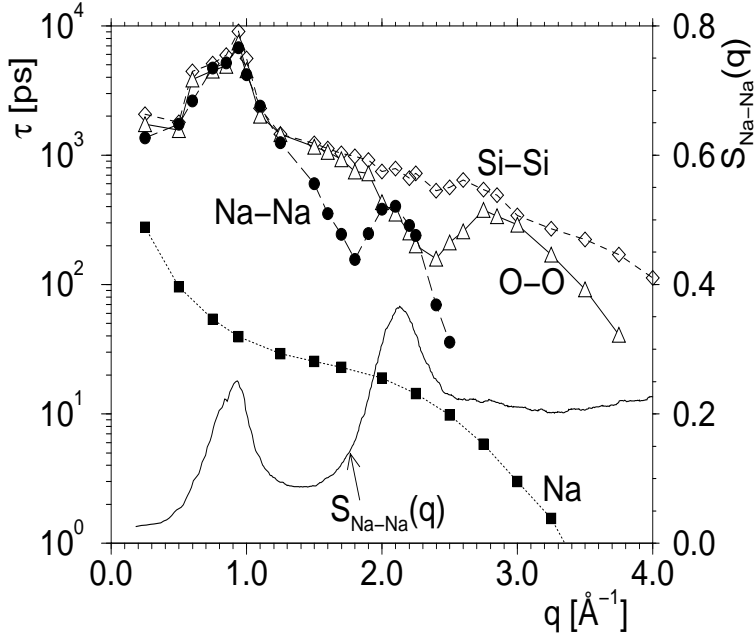,width=10cm,height=8.5cm}
\vspace{0.2cm}
\caption{$\tau$ from $F(q,t)$ for the Si--Si (open diamonds), O--O 
         (triangles), and Na--Na (filled circles) correlations and $\tau_{\rm s}$ from 
	 $F_{{\rm s}}(q,t)$ for Na (filled squares). 
         The labels for the $y$ axis on the right--hand side are for the partial
	 static structure factor $S_{{\rm Na-Na}}(q)$ (bold solid line).
	 }
\label{fig3}
\end{figure}

We address now the question how the sodium ions relax inside the
channels. An appropriate quantity to investigate this point are
time dependent density--density correlation functions, i.e.~the
coherent intermediate scattering function $F(q,t)$ and its self
part, the incoherent intermediate scattering function $F_{{\rm
s}}(q,t)$~\cite{boon}.  In Fig.~\ref{fig2} we show $F(q,t)$ for
the Na--Na correlations (solid lines) as well as $F_{\rm s}(q,t)$
for the sodium atoms (dashed lines) for three different wave-vectors:
$q=0.94$~\AA$^{-1}$, $2.0$~\AA$^{-1}$, and $3.5$~\AA$^{-1}$. From this
figure we infer immediately a surprising result: At $q=0.94$~\AA$^{-1}$,
i.e.~at the characteristic $q$ value of the sodium channel structure,
$F(q,t)$ decays on a time scale which is two orders of magnitude larger
than the one for $F_{\rm s}(q,t)$. Such a strong difference cannot
be explained by a de Gennes narrowing argument~\cite{boon}. Instead
this result can be rationalized by the idea that the sodium atoms move
quickly between preferential sites, since this type of motion gives
rise to a fast decorrelation of the incoherent function whereas it does
not affect the coherent one. Only on the time scale of the relaxation of
the SiO$_2$ matrix also the coherent function starts to decay (see also
Fig.~\ref{fig3}, discussed below). Note that the slow decay of $F(q,t)$
is found only for wave--vectors around 0.95~\AA$^{-1}$. For different $q$
the function decays significantly faster as can be seen from the other
curves shown in Fig.~\ref{fig2}.

In order to quantify the typical relaxation time of the two mentioned
processes we define the $\alpha-$relaxation times $\tau$ and $\tau_{\rm
s}$ as the times at which $F(q,t)$ and $F_s(q,t)$, respectively, have
decayed to 0.1. In Fig.~\ref{fig3} we show $\tau(q)$ for the Si--Si and
O--O correlation, i.e. the relaxation time of the matrix for wave--vector
$q$ (open symbols). Also included is $\tau(q)$ for the Na--Na correlation
and we see that for wave--vectors larger than $\approx1.5$\AA$^{-1}$ it is smaller than the
characteristic relaxation time of the network. In contrast to this we find
that for $q \lesssim q_1$ the relaxation time for Na--Na is close to the
one of the matrix. 
Moreover we see that $\tau(q)$ for the Na--Na correlation
is in phase with the corresponding partial structure factor $S_{{\rm Na-Na}}(q)$
which is also shown in the figure.
Furthermore we recognize from this figure that the
relaxation time of {\it all} coherent correlators is largest at $q_1$,
in agreement with the discussion of Fig.~\ref{fig2}. Also included in
the figure is $\tau_s$ for the sodium atoms and we see that it is indeed
much smaller than the one of $F(q,t)$, which shows that there are indeed
two different processes.

Having discussed the relaxation dynamics of the SiO$_2$ matrix and
the sodium atoms on a qualitative level, we proceed now to investigate
this dynamics in more detail.  In this context it is of interest that
a calculation of Bosse and Kaneko within the mode coupling theory
(MCT)~\cite{mct} found that in a binary system of hard spheres with
large size ratio the coherent and incoherent scattering functions decay
on very different time scales~\cite{bosse97}. Furthermore it was found
that the diffusion constant of the small particles is orders of magnitude
larger than the typical relaxation time of the larger particles. Thus
qualitatively the behavior of that system was very similar to the present
one, with the quasi frozen SiO$_2$ matrix and the sodium atoms playing
the role of the large and small particles, respectively. This similarity
also suggests that MCT might be able to give a reliable {\it qualitative}
description of the relaxation dynamics of the present system.  Indeed we
show elsewhere~\cite{horbach01a} that the incoherent and coherent dynamics
of silicon and oxygen in NS2 can be nicely described by the universal
predictions of MCT. In particular the
time--temperature superposition principle for the $\alpha-$relaxation
regime holds and the late $\beta-$relaxation regime, i.e.~the time window
in which a correlation function $\Phi_q(t)$ starts to fall below the plateau at
intermediate times, can be described by a von Schweidler law,

\begin{equation}
   \Phi_q(t) = f_q^{{\rm c}} + h_q t^b
   \label{eq2}
\end{equation}
where $f_q^{{\rm c}}$ is the height of the mentioned plateau
in $\Phi_q(t)$ (=$F(q,t)$ or $F_{{\rm s}}(q,t)$), and $h_q$ is a
prefactor. According to MCT the exponent $b$ in Eq.~(\ref{eq2}) should be
a system universal constant, i.e.~be independent of the correlator or
$q$, and indeed all our time correlation functions can be fitted with
Eq.~(\ref{eq2}) by fixing the exponent $b$ to $0.47$.

\begin{figure}
\psfig{file=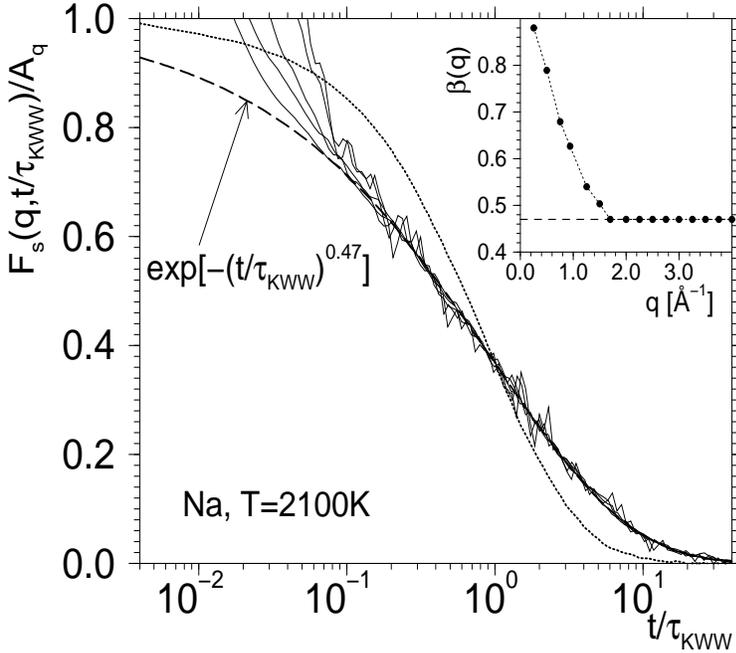,width=10cm,height=8.5cm}
\vspace{0.2cm}
\caption{The solid lines show $F_{{\rm s}}(q,t/\tau_{{\rm KWW}}(q))/A_q$ 
         for $q=1.7$~\AA$^{-1}$, $2.0$~\AA$^{-1}$, $2.75$~\AA$^{-1}$, $3.5$~\AA$^{-1}$,
	 and $4.0$~\AA$^{-1}$ (from left to right). 
         (See text for the determination of $A_q$ and
	 $\tau_{{\rm KWW}}$.) The dotted line shows the same quantity
	 for $q=0.5$~\AA$^{-1}$. The dashed line is the  
         Kohlrausch law $\Phi (\hat{t}) = \exp(- (t/\tau_{{\rm KWW}})^{0.47})$. 
	 The inset shows the Kohlrausch exponent $\beta$ as a function of $q$.}
\label{fig4}
\end{figure}

In the following we will relate this result with the relaxation dynamics
of the sodium atoms as measured by $F_{\rm s}(q,t)$. The long time--decay
of these functions can be well described by Kohlrausch laws,

\begin{equation}
  \Phi_q (t) = A_q \exp( - (t / \tau_{{\rm KWW}}(q))^{\beta(q)}) \  .
  \label{eq3}
\end{equation}
The $q$--dependence of the exponent $\beta$ is shown in the inset of
Fig.~\ref{fig4}. Very surprisingly we see that, for $q>1.6$~\AA$^{-1}$,
$\beta$ seems to become independent of $q$ and is around
$\beta_\infty=0.47$. That $\beta$ really becomes independent of $q$
is demonstrated in Fig.~\ref{fig4} where we plot $F_{{\rm s}}(q,t)/A_q$
versus the rescaled time $t/\tau_{{\rm KWW}}$ for $1.7$~\AA$^{-1} \le
q \le 4$~\AA$^{-1}$. (Note that $A_q$ and $\tau_{{\rm KWW}}$ have been
obtained from fits to $F_{{\rm s}}(q,t)$ in which $\beta$
was fixed to $\beta_\infty$.) If $\beta = \beta_\infty$ really holds
the curves for the different $q$ should fall onto a master curve in the
$\alpha-$relaxation regime and they do so for $t/\tau_{{\rm KWW}} > 0.1$.
That this is a nontrivial result is demonstrated by the curve for $q=0.5$ which is
included in the figure as well.

We note that the value of $\beta_\infty$ coincides with the value of
the von Schweidler exponent $b$ from Eq.~(\ref{eq2}). This important
observation can be rationalized by the prediction of Fuchs, obtained
within a MCT calculation, that the Kohlrausch law becomes exact
for $q \to \infty$ in which case $\beta_\infty$ is equal to the von
Schweidler exponent $b$~\cite{fuchs94}. Fuchs has argued that this
result can be interpreted using Levy's generalization of the central
limit theorem~\cite{gnedenko}: For large $q$ the $\alpha-$relaxation of
the correlators is the result of a sum of many independent terms, each
of which follows the von Schweidler law Eq.~(\ref{eq2}).  According to
Levy's central limit theorem this sum is the characteristic function of a stable
distribution and the latter is exactly the Kohlrausch law with exponent
$b$.  For finite values of $q$ the short time asymptote of the 
$\alpha-$relaxation regime is still a von Schweidler law but one has strong
correlations of correlators with different $q$. So one has no longer the
sum of independent processes, and the Kohlrausch law with exponent $b$
is no longer a solution for the $\alpha-$relaxation regime.  In this
sense the theory of the $\alpha-$process in MCT is a generalization of
Levy's central limit theorem.

If we apply this probabilistic interpretation of $F_{{\rm s}}(q,t)$
for sodium we obtain two important results: Firstly Levy's central limit
theorem can already be applied for $q>1.6$~\AA$^{-1}$, i.e.~deviations
become important only on length scales which are larger than the nearest
neighbor distance between sodium atoms. Secondly the $\alpha-$processes
for the self--motion of sodium ions and the relaxation of sodium channels
have --- although appearing on completely different time scales ---
the same short--time behavior given by a von Schweidler law with exponent
$b = 0.47$. So it seems that the fast diffusion of sodium ions through
the channels has its slow counterpart in the relaxation of the matrix,
i.e.~the rearrangement of the channels at long times.

In summary we have found that the sodium trajectories in NS2 form
channels in the SiO$_2$ network, in agreement with recent results for
NS4~\cite{jund01}. The characteristic distance between these channels
is of the order of 6.6~\AA, i.e.~the distance between second nearest
Na--Na or Si--Na neighbors, and is reflected in the static structure
factor as a prepeak at $q=0.95$~\AA$^{-1}$. Structural relaxation in
NS2 is dominated by two processes: the fast sodium diffusion through
the channels and the rearrangement of these channels on the time scale
of the relaxation of the SiO$_2$ matrix. The von Schweidler exponent $b$
for the system can also be found in the incoherent intermediate scattering
function for sodium in that for $q>1.6$~\AA$^{-1}$ its long time decay
is given by a Kohlrausch law with $\beta_\infty = b$. This shows that
the relaxation of channels at long times and the fast sodium diffusion
inside these channels are intimately connected to each other.

Acknowledgments: Part of this work was supported by the DFG through
SFB 261/D1. We thank the RUS for a generous grant in computer time.


%
\newpage

\end{document}